\documentclass[aps,prl,preprint,superscriptaddress,floatfix]{revtex4-1}

\usepackage{graphicx}
\usepackage{amsmath, amsthm, amssymb}
\begin{document}

%\preprint{}

\title{Localized states qualitatively change the response of ecosystems to varying conditions and local disturbances}


\author{Yuval R. Zelnik}
\affiliation{Department of Solar Energy and Environmental Physics, BIDR, Ben-Gurion University of the Negev, Sede Boqer Campus 84990, Israel}
\author{Ehud Meron}
\affiliation{Department of Solar Energy and Environmental Physics, BIDR, Ben-Gurion University of the Negev, Sede Boqer Campus 84990, Israel}
\affiliation{Department of Physics, Ben-Gurion University of the Negev, Beer Sheva, 84105, Israel}
\author{Golan Bel}
\email[]{bel@bgu.ac.il}
\affiliation{Department of Solar Energy and Environmental Physics, BIDR, Ben-Gurion University of the Negev, Sede Boqer Campus 84990, Israel}

%\thanks{}
%\affiliation{}


\date{\today}

\begin{abstract}
The response of dynamical systems to varying conditions and disturbances is a fundamental aspect of their analysis.
In spatially extended systems, particularly in pattern-forming systems, there are many possible responses, including critical transitions, gradual transitions and locally confined responses. 
Here, we use the context of vegetation dynamics in drylands in order to study the response of pattern-forming ecosystems to oscillating precipitation and local disturbances.
We focus on two precipitation ranges, a bistability range of bare soil with a patterned vegetation state, and a bistability range of uniform vegetation with a patterned vegetation state.
In these ranges, there are many different stable states, which allow for both abrupt and gradual transitions between the system states to occur.
We find that large amplitude oscillations of the precipitation rate can lead to a collapse of the vegetation in one range, while in the other range, they result in the convergence to a patterned state with a preferred wavelength. 
In addition, we show that a series of local disturbances results in the collapse of the vegetation in one range, while it drives the system toward fluctuations around a finite average biomass in the other range. 
Moreover, it is shown that under certain conditions, local disturbances can actually increase the overall vegetation density.
These significant differences in the system response are attributed to the existence of localized states in one of the bistability ranges.
\end{abstract}
%\pacs{}
%\keywords{}
\maketitle

\section{Introduction}\label{intro}
The dynamics of ecosystems are often complex and nonlinear. 
This nonlinearity, originating from various feedbacks between the different components of the ecosystem, may result in a multistability of the ecosystem states. 
Changes in environmental conditions and disturbances may drive the system from one stable state to an alternative one. 
These critical transitions or ``regime shifts'' may be either abrupt or gradual \cite{bel2012theo_ecol}. 
Abrupt (critical) transitions are of great concern in many fields of science due to the significant changes and often unexpected outcomes they entail \cite{Gandhi1998,Barnosky2012,Sun2013,Rietkerk2011,Dakos2012,Kefi2014,Cline2014}.
The basic notion of abrupt critical transitions stems from mean field models, describing the dynamics of variables that are uniform across the system.
The dynamics and responses of spatially extended systems, however, are more complicated \cite{Rietkerk2011,Dakos2012,Kefi2014,Cline2014,Fort2009,Fort2013review}. 
In particular, the regime shifts between alternative stable states may be gradual, abrupt or even a combination of the two, appearing as a sequence of local regime shifts that can eventually lead to a transition of the entire system to another stable state.

An excellent case study for studying regime shifts in spatially extended systems can be found in dryland landscapes, where fascinating vegetation patterns have been observed and studied 
\cite{Klausmeier1999science,vonhardenberg2001prl,Rietkerk2002an,sherratt-2005-analysis,Tlidi2008lnp,Rietkerk2004science,manor-2008-facilitation,Lejeune2002pre,Lejeune2004ijqc,HilleRisLambers2001ecology,Gilad2007jtb,Gilad2004prl,borgogno2009gr}. 
The vegetation patterns are driven and maintained by positive feedbacks between local vegetation growth and water transport toward the growth location \cite{Kinast2014prl,Meron2015book}. 
Several pattern-forming feedbacks can be distinguished according to the water transport mechanism: overland water flow, water conduction by laterally extended root systems, soil water diffusion and fog advection \cite{Meron2015book}.
Models describing the dynamics of dryland vegetation differ in the pattern-forming feedbacks they capture, but they all show the same universal sequence of basic vegetation states along the rainfall gradient: 
uniform vegetation, gap patterns, striped patterns, spotted patterns and uniform bare soil. 
In addition, a bistability range of each pair of consecutive basic states may exist \cite{meron2012ecomod,Gowda2014pre}.
The multiplicity of stable states, however, is much higher. 
Each type of basic periodic pattern represents a family of patterns with different wavelengths, e.g., a family of periodic striped patterns, ranging from isolated stripes to dense stripes~\cite{vanderStelt2013nonl_sci,zelnik2013regime}. 
In addition, any bistability range of basic states may give rise to a multiplicity of hybrid states involving domains of one state embedded within a larger domain of another state~\cite{Zelnik2015}.
This multiplicity of stable states allows for different types of state transitions and may consequently lead to a wide array of responses to varying environmental conditions or disturbances.

 Ecosystem responses to changes in global conditions, such as global climate change and climate variability, have attracted much attention in various contexts
\cite{walther2002,marshall2008,parmesan2006,walther2010,Anderson2012,Maclean2011,cramer2001,brown1997,melillo1993,swetnam2010,porporato2004,pounds1999,mcgowan1998climate,yizhaqwrr2014}. 
The responses of ecosystems to disturbances, be they natural, such as fires and bark beetles, or anthropogenic, 
such as cattle grazing and clear cutting, have also been the subject of much research \cite{Turner2003,Goetz2007,abdelnour2011,white2001search,mcmillan2011}. 
Several recent model studies of dryland vegetation have addressed the response of vegetation patterns to such influences, 
that is, to large-scale environmental changes that encompass the whole ecosystem (hereafter ``global'' changes) and to confined disturbances that directly affect limited parts of the ecosystem.
It was shown that in the bistability range of bare soil and patterned vegetation, the system can respond to global changes, such as a gradual precipitation decrease, 
by changing the wavelength of the pattern \cite{zelnik2013regime,vanderStelt2013nonl_sci,Sherratt20138,Sherratt2014,Siteur201481}. 
The Busse balloon, which presents the range of stable wavelengths versus the bifurcation parameter, provides an insight into this type of response \cite{vanderStelt2013nonl_sci}. 
When the bifurcation parameter is adiabatically changed to values outside the stability range of the current wavelength, a transition to another wavelength within the Busse balloon takes place. 
The new wavelength that is chosen by the system depends on the rate of precipitation decrease and the noise level~\cite{Siteur201481}.
The effect of system parameters varying periodically or intermittently in time has also been studied \cite{guttal2007self,Kletter2009jtb,Sheffer2011jtb,zhao2014rich,Gandhi2015prl}, 
but not in the context of global state transitions or regime shifts.

To the best of our knowledge, a thorough examination of transitions due to periodic forcing or local disturbances has not been performed for models describing the dynamics of dryland vegetation.
Furthermore, all of the studies mentioned above have focused on the bistability range of bare soil with patterned vegetation, and have not looked at other bistability (or multistability) ranges of the system.
Here, we study the response of pattern-forming systems to local disturbances and temporal changes in the control parameter 
using a simple pattern-forming model that describes vegetation dynamics in dryland ecosystems and exhibits various patterns in different precipitation regimes.
We focus on the responses of the system in two precipitation ranges in which there is a bistability between uniform and patterned states.
The responses of the system in the bistability range of bare soil (uniform zero-biomass state) and patterns, and in the bistability range of uniform vegetation (uniform nonzero-biomass state) and patterns are studied and compared.
%~~~~~~~~~~~~~~~~~~~~~~~~~~~~~~~~~~~~~~~~~~~~~~~~~~~~~~~~~~~~~~~~~~~~~~~~~~~~~~~~~~~~~~~~~~~~~~~~~~~~~~~~~~~~~~~~~~~~~~~~~
%~~~~~~~~~~~~~~~~~~~~~~~~~~~~~~~~~~~~~~~~~~~~~~~~~~~~~~~~~~~~~~~~~~~~~~~~~~~~~~~~~~~~~~~~~~~~~~~~~~~~~~~~~~~~~~~~~~~~~~~~~
\section{Model and Methods}\label{sec:model}
We studied a relatively simple model that describes the spatio-temporal distributions of soil water and aboveground biomass.
However, the results are not limited to this specific model, as detailed in the appendix.
The model is a simplified version~\cite{zelnik2013regime} of the model that was introduced by Gilad et al. \cite{Gilad2004prl,Gilad2007jtb}.
It applies to plants that have confined root zones in the lateral directions and to landscapes in which there is no infiltration contrast between vegetated domains and bare-soil domains, e.g., soils without significant crust.
This simplified model captures a single pattern-forming feedback, the ``uptake-diffusion'' feedback~\cite{Kinast2014prl}. 
This feedback represents the increased water uptake rate by denser vegetation, thereby reducing the soil water density in its neighborhood, 
and the fast transport of soil water from domains with sparse vegetation (and therefore, higher soil water density) toward domains with denser vegetation (and therefore, lower soil water density).
This mechanism is relevant to dryland ecosystems with large soil water diffusivity, such as landscapes with sandy soil, and plants whose water uptake rate has a nonlinear dependence on the biomass density. 
For example, the ecosystem of fairy circles in Namibia has been shown to be a system where this mechanism is prominent~\cite{Zelnik2015}.
Other pattern-forming feedbacks exist, such as the infiltration feedback~\cite{Rietkerk2011,meron2012ecomod,Kinast2014prl}, which occurs in regions with a high infiltration contrast 
so that denser vegetation patches act as water sinks, and the root-augmentation feedback, which describes the lateral growth of the plants' root systems as biomass density grows~\cite{Gilad2007jtb,meron2012ecomod}. 
For the purpose of this study, the simplified model is detailed enough to capture the distinct ranges of bistability of bare-soil and patterned states and of uniform vegetation and patterned states.
The equations describing the dynamics of the aboveground biomass ($B$) and the soil water ($W$) areal densities are:

\begin{equation}
\partial_T B = \Lambda WB(1 + E B)^2(1 - B/K) - M B + D_B\nabla^2B \,, \label{SGeqb}
\end{equation}
\begin{equation}
\partial_T W = P - N W (1-R B/K) - \Gamma WB(1 + E B)^2  + D_W\nabla^2 W\,. \label{SGeqw}
\end{equation}
In Eq. \eqref{SGeqb}, $\Lambda$ is the biomass growth rate coefficient, $E$ is a measure for the root-to-shoot ratio, which characterizes the positive feedback of the biomass on the soil water uptake rate,
$K$ is the maximum standing biomass, $M$ is the mortality rate, and $D_B$ represents the rate of seed dispersal or clonal growth. 
In Eq. \eqref{SGeqw}, $P$ is the precipitation rate, $N$ is the evaporation rate, $R$ is a dimensionless factor representing a reduction of the evaporation rate due to shading, 
$\Gamma$ is the water uptake rate coefficient, and $D_W$ is the effective soil water diffusivity in the lateral ($X,Y$) directions.
The ``uptake-diffusion'' feedback is characterized by the parameters $E$ (characterizing the growth of the water uptake rate with the biomass density) and $D_W$ (the soil water diffusivity) \cite{Kinast2014prl}. 
A dimensionless form of the model is obtained by rescaling the state variables $B,W$ and the space and time coordinates as follows:

\begin{equation}
    b = \frac{B}{K}; w = \frac{W\Lambda}{K \Gamma}; t = M T; x = X\sqrt{M/D_B}\,.
\end{equation}
In terms of these dimensionless quantities, the model reads:

\begin{equation}
\partial_t b = \lambda w b(1 + \eta b)^2(1 - b) - b + \nabla^2b \,, \label{ndSGeqb}
\end{equation}
\begin{equation}
\partial_t w = p - \nu w (1-\rho b) - \lambda w b(1 + \eta b)^2  + \delta_w\nabla^2 w\,. \label{ndSGeqw}
\end{equation}
The dimensionless parameters are related to their dimensional counterparts by the following relations:

\begin{equation}
   \nu = \frac{N}{M};   \eta = E K;  \lambda = \frac{K \Gamma}{M};  p = \frac{\Lambda P}{K \Gamma M} ;   \rho = R ;   \delta_w = \frac{D_W}{D_B} \,.
\end{equation}
In what follows, all the quantities presented will be dimensionless. The bifurcation parameter is set to be the precipitation rate, $p$. 
The other parameters of the model were set to the following values: $\nu=2$;$\eta=6$;$\lambda=2$;$\rho=0.2$; and $\delta_w=1000$.
These parameters were chosen to have realistic values in dryland ecosystems \cite{Zelnik2015}.
For simplicity, we focused on the case of one spatial dimension.
The different states of the system, used to create the bifurcation diagrams, were calculated using numerical continuation with the AUTO software \cite{doedel1981auto}.
The stability of these solutions was tested numerically using a linear stability analysis with periodic boundary conditions in a system of size $\sim 250$ 
(the exact size was set to be an integer number of wavelengths in order to allow for periodic boundary conditions).
The numerical integration in time was done using a pseudo-spectral method, with periodic boundary conditions.
We note that the size of the periodic patterns, in a dimensional form, is in the range of $0.1-20$ meters. This range corresponds to the dryland vegetation patterns in different ecosystems in the world~\cite{Deblauwe2008geb}.

The responses of the system to two types of environmental changes were studied. The first type, periodic changes in the precipitation rate, 
was implemented by modulating the precipitation parameter, $p$, sinusoidally over time, with $p(t) = p_0 + A\left(t\right)$, where $ A\left(t\right)= A sin (2\pi t/T_0)$.
This perturbation form is general enough to capture the phenomenon we wish to demonstrate, while other more intricate forms may involve unnecessary complications.
The baseline precipitation value, $p_0$, is the original value of $p$ for which a steady state was reached. 
The modulation period was fixed at $T_0= 120$, and the modulation amplitude was varied within the range $0.04\le A\le 0.16$.
This period amounts to roughly $5-100$ years in dimensional units (depending on the specific choice of dimensionalization), a timescale consistent with observed droughts~\cite{currie1981evidence}.
For the purpose of time integration, the modulation period, $T_0$, was divided into $50$ parts, and for each time segment, an integration over a time of $t=T_0/50$ was performed with a constant value of $p$, 
corresponding to the relevant phase in the beginning of the segment. 
We verified that this choice of discretization of the periodic signal was sufficient; namely, the results of a finer discretization were the same.
The second type of environmental change that we studied represented local disturbances, which was implemented by removing the vegetation (setting $b=0$) in small domains, each of a size $L_0=50$.
The position of each disturbance was chosen from a random uniform distribution, and between sequential disturbances, the system was integrated forward in time for a period of $T_1=100$, 
which is long enough for the system to converge to a new steady state (i.e., a state that shows no significant dynamics over a long period). 
We confirmed that integrating for a longer period after each disturbance did not affect the results.
For clarity, we point out that in studying the response to local disturbances, the precipitation rate was constant.

%~~~~~~~~~~~~~~~~~~~~~~~~~~~~~~~~~~~~~~~~~~~~~~~~~~~~~~~~~~~~~~~~~~~~~~~~~~~~~~~~~~~~~~~~~~~~~~~~~~~~~~~~~~~~~~~~~~~~~~~~~
%~~~~~~~~~~~~~~~~~~~~~~~~~~~~~~~~~~~~~~~~~~~~~~~~~~~~~~~~~~~~~~~~~~~~~~~~~~~~~~~~~~~~~~~~~~~~~~~~~~~~~~~~~~~~~~~~~~~~~~~~~
\section{Bifurcation diagram for constant conditions}
\begin{figure}
    %\begin{center}
    \includegraphics[width=0.95\linewidth]{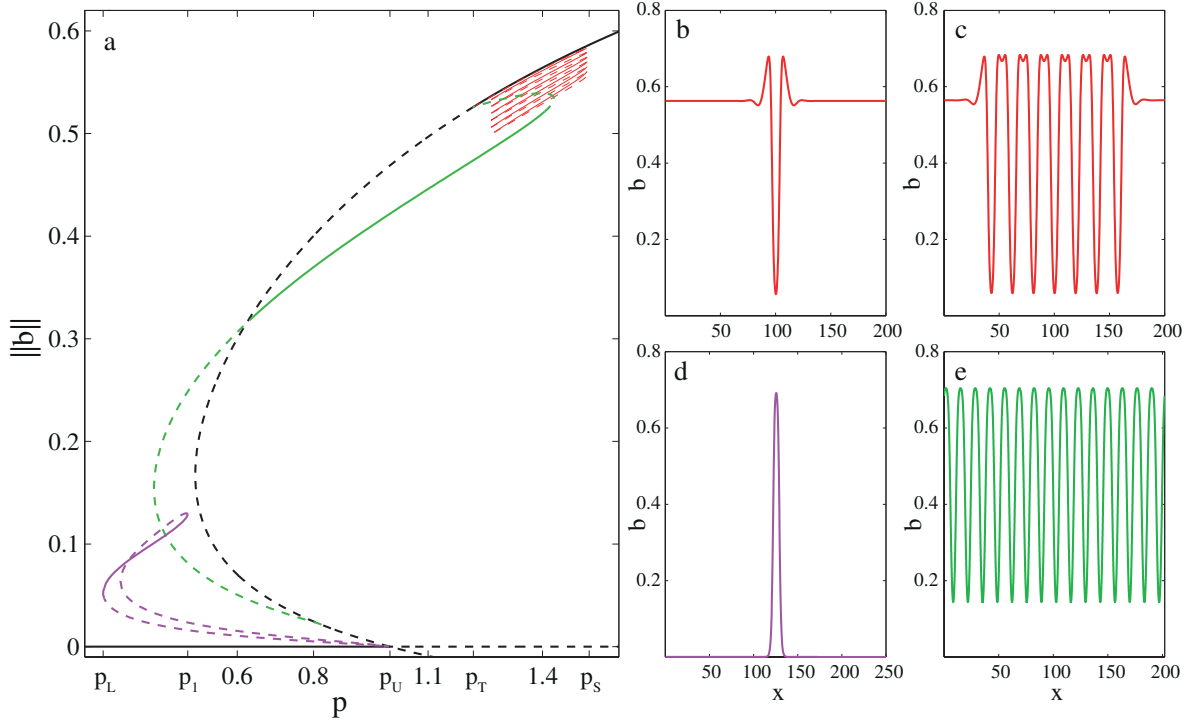}
  %\end{center}
    \caption{\textbf{(a)} A bifurcation diagram of the model in one spatial dimension, showing the two ranges of bistability between uniform states and patterned states. 
  The vertical axis denotes the L2 norm of the biomass density while the horizontal axis denotes the precipitation rate. Solid (dashed) curves denote stable (unstable) solutions. 
  The shown steady states include the two uniform states of the system (black curves, corresponding to the bare-soil ($b=0$) and uniform-vegetation states), the Turing patterned state (green curve), 
  a patterned state with a wavelength equal to the size of the system (magenta curve) and hybrid solutions (red curve).
The spatial distributions of the biomass of different solutions are shown in panels \textbf{b-e}.
\textbf{(b)} Hybrid state of a single hole in a uniform vegetation domain.
\textbf{(c)} Hybrid state of several holes in a uniform vegetation domain.
\textbf{(d)} Long-wavelength periodic solution (a single peak in the domain $L=250$ used).
\textbf{(e)} Short-wavelength periodic solution. \label{fig:BifGen}}
\end{figure}

We began our analysis by looking at the different states of the system under a constant precipitation rate and in the absence of disturbances.
The various possible states of the system for different values of the precipitation rate, $p$, are shown in the bifurcation diagram in Fig. \ref{fig:BifGen}.
The system has two uniform states, the bare-soil state ($b=0$) that exists for all values of $p$, and a uniform-vegetation state ($b>0$) that exists for high enough values of $p$.
The bare-soil state is stable for low values of $p$, and loses its stability at $p=p_U$ in a uniform (zero wavenumber) stationary instability~\cite{Meron2015book}. 
At this point, it crosses an unstable branch that describes uniform vegetation. This solution branch merges with another branch, representing denser uniform vegetation, at a fold bifurcation occurring at a lower value of $p$. 
The latter still describes an unstable state, but the instability is due to the growth of non-uniform perturbations. 
The uniform-vegetation state becomes stable only for precipitation rates higher than $p_T$, which designates a non-uniform stationary (Turing) instability.

\begin{figure}
  %\begin{center}
 \includegraphics[width=0.95\linewidth]{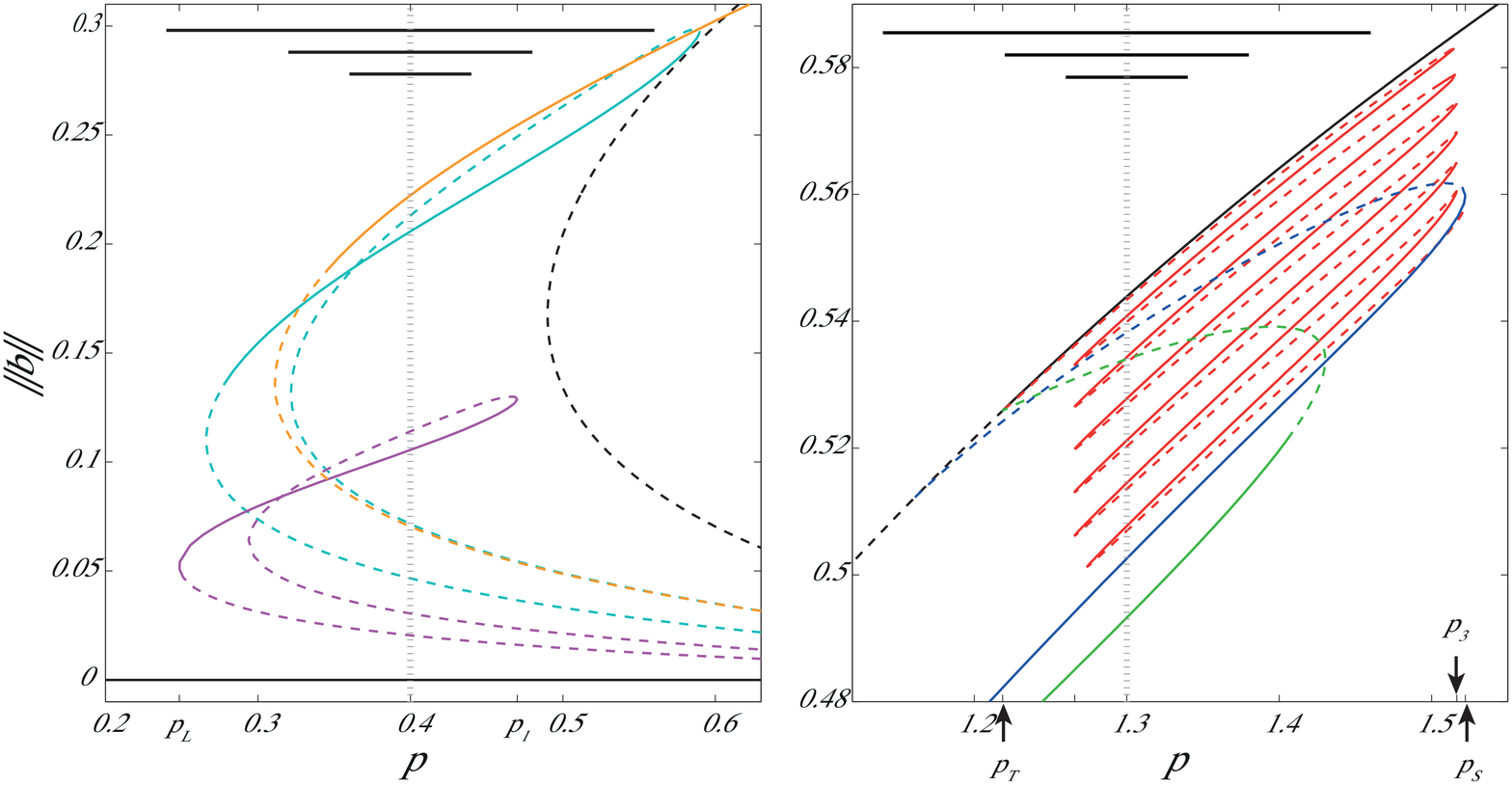}
  %\end{center}
   \caption{Close up of two bistability ranges in the bifurcation diagram shown in Fig \ref{fig:BifGen}.
 \textbf{(a)} Bistability of the bare-soil solution (black solid line) and of periodic solutions (colored curves), $p_L < p < p_U$ (only part of the bistability range is shown, see Fig. \ref{fig:BifGen}).
 The orange and cyan curves show patterned states with wavelengths $25$ and $50$, 
 respectively, and the purple curve shows the patterned state with a wavelength that is equal to the system size (corresponding to a single peak and wavelength $250$).
 \textbf{(b)} Bistability range of the uniform-vegetation solution (black curve) and of periodic solutions (green and blue curves), $p_T < p < p_S$. 
 The green curve shows the periodic solution emanating from the Turing point $p_T$, and the blue curve shows the periodic solution (with wavelength $20$) that extends to the boundary, $p=p_S$, of the bistability range. 
 Also shown are hybrid solutions (red curve).
 The horizontal black lines on top indicate the ranges of the precipitation rate spanned by the sinusoidal modulations ($A=0.04,0.08$ and $0.16$, respectively), with the dotted vertical gray lines indicating the average value of $p$ for these modulations.\label{fig:BifSub}}
\end{figure}

The periodic solutions represent periodic vegetation patterns. Many solutions of this kind exist, with two representative examples shown in Fig. \ref{fig:BifGen}.
The first (shown in green in Fig. \ref{fig:BifGen}a,e) is the periodic-solution branch that emanates from the Turing bifurcation point, $p_T$, and continues all the way down to the unstable uniform-vegetation solution.
The second (shown in magenta in Fig. \ref{fig:BifGen}a,d) is a single-peak solution, that is, a periodic solution with a wavelength equal to the system size and thus much larger than the wavelength of the first solution. 
It emanates from and terminates in the small-amplitude uniform-vegetation solution close to the bifurcation point at $p=p_U$, and describes an isolated vegetation spot in otherwise bare soil.
In addition to periodic solutions, non-periodic solutions also exist (shown in red in Fig. \ref{fig:BifGen}a,b,c) that describe hybrid states, i.e., confined domains of a periodic pattern in an otherwise uniform vegetation domain.  
The first solution of this kind emanates from the uniform-vegetation solution at the Turing bifurcation point (the first periodic solution described above emanates from the same point), and describes a single gap in uniform vegetation.
The hybrid-solution branch snakes down and terminates in a different periodic-solution branch (the blue curve in Fig. \ref{fig:BifSub}b). 
The hybrid solutions differ in the size of the patterned domain, which increases through the appearance of a new pair of gaps (one gap on each side of the domain) at each step in the descent toward the periodic-solution branch. 
The hybrid solutions shown in Fig. \ref{fig:BifGen} describe confined domains with odd numbers of gaps.
A corresponding set of hybrid solutions with even numbers of gaps also exist (not shown).
The stability range of an isolated vegetation spot (single-peak solution) lies within a larger bistability range of bare soil and periodic patterns, shown in detail in Fig. \ref{fig:BifSub}a,
while the stability range of the different hybrid states lies within a larger bistability range of uniform vegetation and periodic patterns, shown in detail in Fig. \ref{fig:BifSub}b.

The bistability range of the bare-soil solution and the periodic solutions, partly shown in Fig. \ref{fig:BifSub}a, lies between the fold bifurcation at $p=p_L$, 
below which the single-peak solution ceases to exist, and $p=p_U$ above which the bare-soil solution loses stability.
The fold bifurcation at $p=p_1$, which designates the upper limit of the single-peak existence range, separates the bistability range into two regimes, where distinct dynamical behaviors were found.
For $p>p_1$, isolated peaks of biomass were found to split into two or more peaks, while for $p_L<p<p_1$, isolated peaks remained stable.
This peak-splitting or ``self-replication'' behavior has been found in chemical reactions \cite{lee1995lamellar}, and in studies of the Gray-Scott model \cite{doelman2000slowly}, 
which is closely related to the vegetation model introduced by Klausmeier \cite{Klausmeier1999science}.
The significance of this finding to our study is that in the range $p_L < p < p_1$, the system tends to conserve the number of peaks in the system, since peak splitting does not occur.

The second bistability range of uniform vegetation and periodic patterns, shown in Fig. \ref{fig:BifSub}b,
lies between $p=p_T$, the Turing instability point of the uniform-vegetation solution, and $p=p_S$, 
where the periodic solution that extends to the highest $p$ values (shown in blue in Fig. \ref{fig:BifSub}b) disappears in a fold bifurcation.
Within this range lies the snaking range, $p_2\le p\le p_3$, where the hybrid solutions are stable.

%~~~~~~~~~~~~~~~~~~~~~~~~~~~~~~~~~~~~~~~~~~~~~~~~~~~~~~~~~~~~~~~~~~~~~~~~~~~~~~~~~~~~~~~~~~~~~~~~~~~~~~~~~~~~~~~~~~~~~~~~~
%~~~~~~~~~~~~~~~~~~~~~~~~~~~~~~~~~~~~~~~~~~~~~~~~~~~~~~~~~~~~~~~~~~~~~~~~~~~~~~~~~~~~~~~~~~~~~~~~~~~~~~~~~~~~~~~~~~~~~~~~~
\section{Responses to Oscillating Precipitation} \label{perper}

\begin{figure}
  %\begin{center}
 \includegraphics[width=0.95\linewidth]{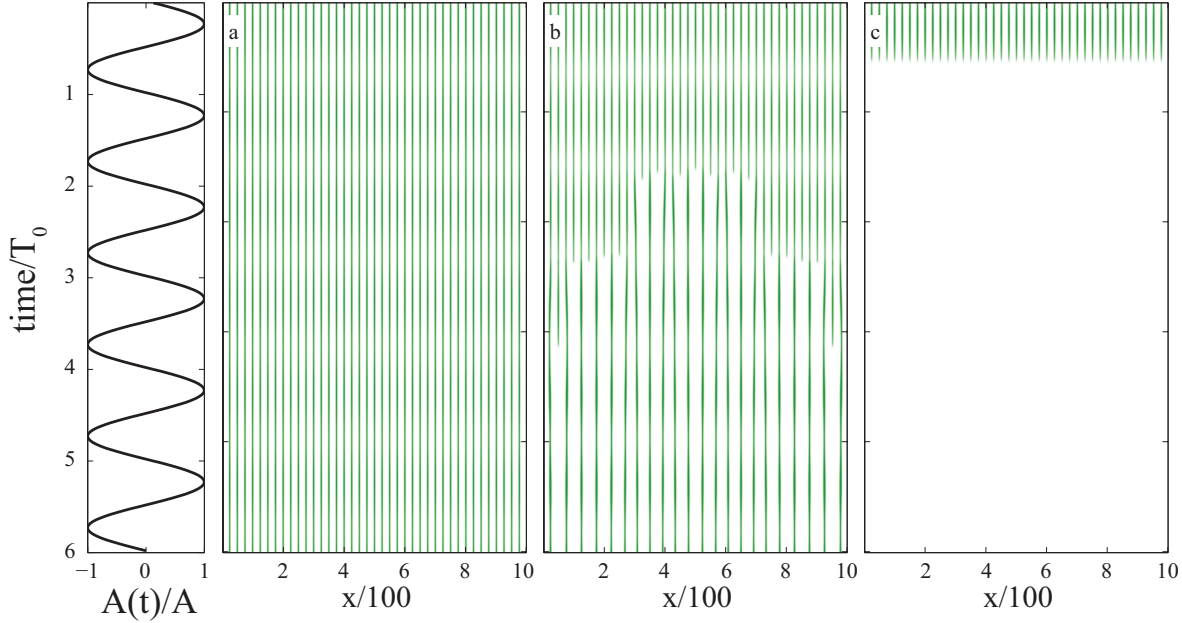}
  %\end{center}
   \caption{The response of the system to periodic modulations of the precipitation rate in the bistability range of bare-soil and patterned states. 
 Shown are space-time plots of the biomass variable (darker shades denote higher biomass density) for three modulation amplitudes: small, $A=0.04$ (panel \textbf{a}), moderate, $A=0.08$ (panel \textbf{b}), and large, $A=0.16$ (panel \textbf{c}). 
 In all three cases, the initial state of the system is the same--a stable periodic pattern solution ($WL=25$). 
 The weak modulation leaves the initial pattern unchanged apart from amplitude oscillations that follow the modulations of the precipitation rate. 
 The moderate modulation induces a transition to a periodic pattern with a wavelength about twice as large as the initial one. 
 The transition largely occurs within the first few periods of the modulated precipitation, as can be inferred from the leftmost panel that shows a plot of $[p(t)-p_0]/A=\sin{(2\pi t/T_0)}$. 
 The strong modulation induces an abrupt transition to the bare-soil state within the first periods of the modulated precipitation. \label{fig:PerPerEvo1}}
\end{figure}

Following the analysis of uniform, periodic and hybrid solutions at constant precipitation, as depicted by the bifurcation diagram in Fig. \ref{fig:BifGen},
we turned to study the possible responses of the system to time-periodic precipitation (as described in Section \ref{sec:model}).
Beginning with the bistability range of the patterned states with the bare-soil state, we used as an initial condition a periodic solution with wavelength $WL=25$. 
The initial condition is typical in the sense that starting from a random initial condition and allowing the system to evolve under constant precipitation 
(equal to the mean value, $p_0$, of the oscillating precipitation that we used) will cause it to converge into a patterned state with that wavelength.
The mean precipitation rate was set to $p_0=0.4$, and the responses to different modulation amplitudes were studied numerically.
As expected, small-amplitude modulations of $p$ ($A=0.04$) have no significant effect on the system, as seen in Fig. \ref{fig:PerPerEvo1}a.
Increasing the modulation amplitude ($A=0.08$), so that a significant period of time is spent outside the stability range of the initial periodic solution, induces a transition to another state with a larger wavelength, shown in Fig. \ref{fig:PerPerEvo1}b.
Since the stability range of this new state is fully within the range spanned by the modulated precipitation, the wavelength does not change further (although the amplitude of the vegetation pattern does change in time according to the value of $p$).
Increasing the modulation amplitude further ($A=0.16$), such that over long enough periods, $p$ lies outside the bistability range, $p<p_L$, can result in a quick collapse to the bare-soil state through an abrupt critical transition, as shown in Fig. \ref{fig:PerPerEvo1}c.

\begin{figure}
    %\begin{center}
  \includegraphics[width=0.95\linewidth]{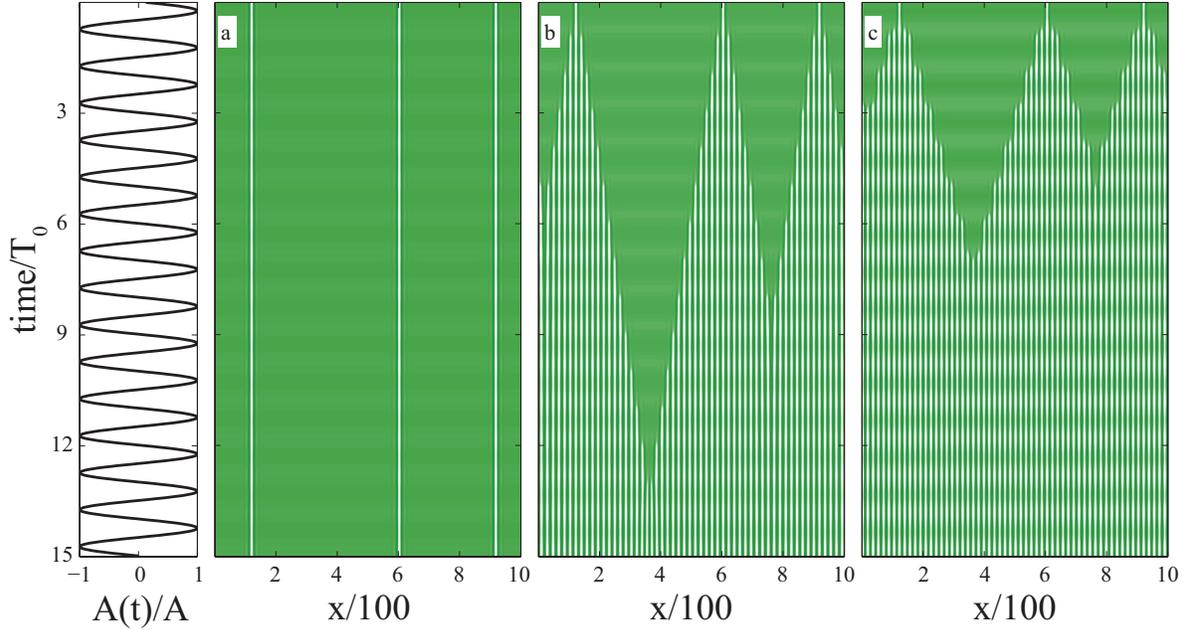}
  %\end{center}
  \caption{The response of the system to periodic modulations of the precipitation rate in the bistability range of uniform-vegetation and patterned states. 
 Shown are space-time plots of the biomass variable (darker shades denote higher biomass density) for three modulation amplitudes: small, $A=0.04$ (panel \textbf{a}), moderate, $A=0.08$ (panel \textbf{b}), and large, $A=0.16$ (panel \textbf{c}). 
 The leftmost panel shows a plot of $A(t)/A=\sin{(2\pi t/T_0)}$.
 In all three cases, the initial state of the system is the same--a stable hybrid solution made up of three separated gaps embedded in a uniform vegetation domain. 
 The weak modulation leaves the initial state unchanged apart from amplitude oscillations that follow the modulation of the precipitation rate. 
 The moderate modulation induces a transition to the patterned state. The transition evolves gradually through the appearance of a new pair of gaps at every period of the modulated precipitation. 
 The strong modulation induces a faster yet  gradual transition that involves the appearance of several pairs of gaps at every period of the modulated precipitation.\label{fig:PerPerEvo2}}
\end{figure}

In studying the possible responses of the system in the second bistability range of patterned and uniform-vegetation states, we used a hybrid state as an initial condition.
In this bistability range, we set the mean precipitation rate to $p_0=1.3$.
Small-amplitude modulations of the precipitation rate ($A=0.04$) have no discernible effect, and the system stays much the same, as can be seen in Fig. \ref{fig:PerPerEvo2}a.
For a larger amplitude ($A=0.08$), where $p$ assumes values outside the snaking range for sufficiently long periods, a gradual transition from the hybrid state to a patterned state, as shown in Fig. \ref{fig:PerPerEvo2}b, occurs. 
With each period of the oscillations, a new pair of gaps appears, with one new gap on each side of the patterned domain. 
The sequential appearance of new gaps results in the expansion of the patterned domain across the whole system and the convergence to a periodic patterned state. 
Since the range of the modulated precipitation rate lies entirely within the stability range of the periodic patterned state, only oscillations of the pattern amplitude occur after the convergence.
Using an even larger amplitude ($A=0.16$) that takes $p$ outside the stability range of the uniform-vegetation state speeds up the convergence to the patterned state, 
by allowing for several new pairs of gaps to appear with every cycle of the modulated precipitation.
This transition, shown in Fig. \ref{fig:PerPerEvo2}c, is still gradual and converges to the same final state (with the only difference being the magnitude of the patterned state’s amplitude oscillations).

The response of the system in the bistability range of uniform-vegetation and pattern solutions is quite different from the response in the bistability range of bare-soil and pattern solutions.
In the former, the transitions are gradual and, in large systems, may take a very long time to complete. 
Moreover, the final state is independent of the amplitude of the precipitation modulation, provided it is large enough to initiate a transition. 
By contrast, in the latter bistability range, the transitions occur over short time scales and do not necessarily result in the same final state; 
moderate amplitudes drive the system from the initial periodic patterned state to various other periodic patterned states, whereas large amplitudes induce transitions to the bare-soil state.

\begin{figure}
    %\begin{center}
  \includegraphics[width=0.95\linewidth]{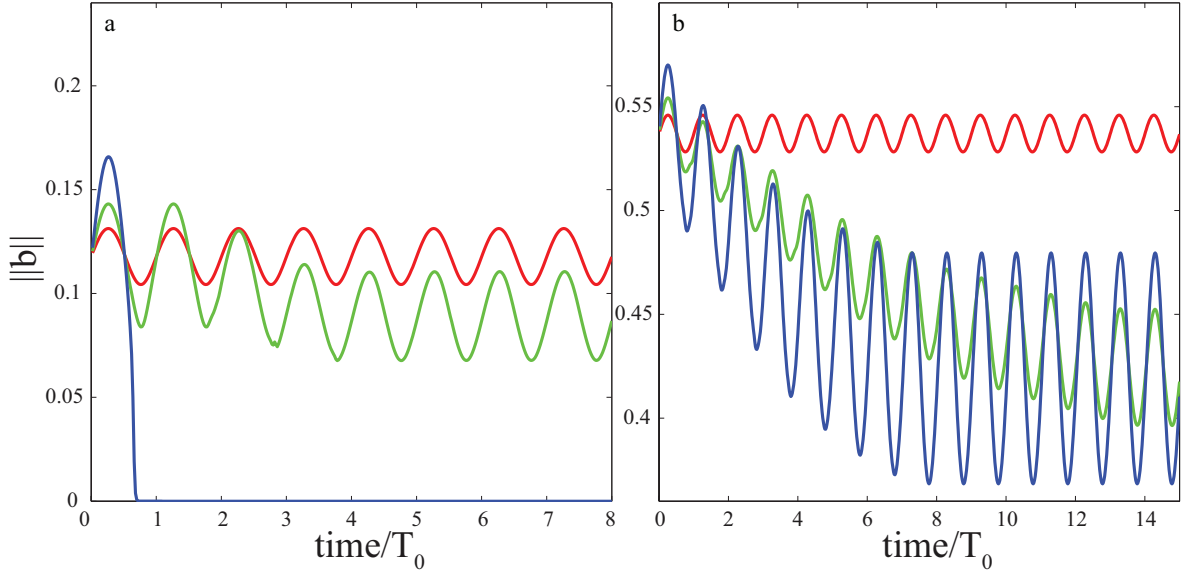}
  %\end{center}
\caption{The spatial average of the biomass vs. time for the scenario shown in Fig. \ref{fig:PerPerEvo1} (panel \textbf{a}) and for the scenario shown in Fig. \ref{fig:PerPerEvo2} (panel \textbf{b}). 
  The red, green, and blue colors denote weak, moderate, and strong modulations of the precipitation rate, respectively. While in the bistability range of bare-soil and patterned states, 
  the biomass changes occur on short time scales and can be abrupt (panel \textbf{a}), in the bistability range of uniform-vegetation and patterned states, the biomass changes are gradual and longer.\label{fig:PerPerBio}}
  \end{figure}

The changes in the average biomass for the scenarios presented in Figs. \ref{fig:PerPerEvo1} and \ref{fig:PerPerEvo2} are shown in Fig. \ref{fig:PerPerBio}a,b, respectively.
The average biomass changes periodically, following the modulation of the precipitation, and in both precipitation ranges, modulations that are strong enough bring the system to lower biomass levels.
In the bistability range of bare soil and periodic patterns, the biomass changes occur quickly, within the first few modulation periods, and can be abrupt for sufficiently large modulations.
On the other hand, in the bistability range of uniform vegetation and periodic patterns, the changes are much more gradual, and the final biomass oscillates around a similar mean value. 
The duration of the transition to the final state, however, may strongly depend on the modulation strength.

%~~~~~~~~~~~~~~~~~~~~~~~~~~~~~~~~~~~~~~~~~~~~~~~~~~~~~~~~~~~~~~~~~~~~~~~~~~~~~~~~~~~~~~~~~~~~~~~~~~~~~~~~~~~~~~~~~~~~~~~~~
%~~~~~~~~~~~~~~~~~~~~~~~~~~~~~~~~~~~~~~~~~~~~~~~~~~~~~~~~~~~~~~~~~~~~~~~~~~~~~~~~~~~~~~~~~~~~~~~~~~~~~~~~~~~~~~~~~~~~~~~~~
\section{Responses to Random Local Disturbances} \label{locdis}
\begin{figure}
      %\begin{center}
  \includegraphics[width=0.95\linewidth]{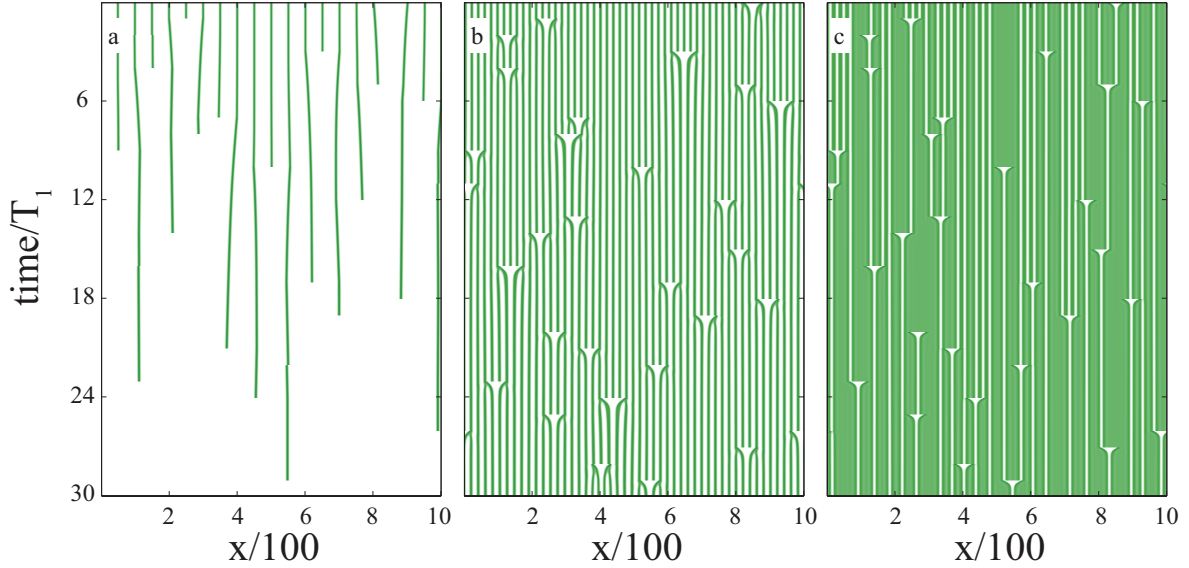}
  %\end{center}
  \caption{The response of the system to periodic local biomass-removal disturbances. Shown are space-time plots of the biomass variable in three distinct precipitation ranges in which different initial conditions were used. 
 \textbf{(a)} $p_L<p=0.4<p_1$: Starting with an initial condition of a periodic solution with a wavelength of $WL=50$, the number of spots (peaks) drops over time, with no new spots being formed. Finally, a bare-soil state is reached. 
 \textbf{(b)} $p_1<p=0.9<p_T$: Starting with an initial condition of a periodic solution with a wavelength of $WL=25$, the system reacts quickly to each disturbance, refilling the bare space by spot splitting. 
 \textbf{(c)} $p_T<p=1.3<p_S$; Starting with an initial condition of a hybrid solution, each disturbance results in a single hole of a certain length.
 If previously the area had several holes, then the overall number of holes is lowered, while if the disturbed domain was vegetated, the number of holes increases. 
 Therefore, after a long enough time, the average number of holes, which determines the overall biomass due to the regular size of the holes, no longer changes significantly.\label{fig:LocDisEvo}
}
\end{figure}

\begin{figure}
    %\begin{center}
  \includegraphics[width=0.95\linewidth]{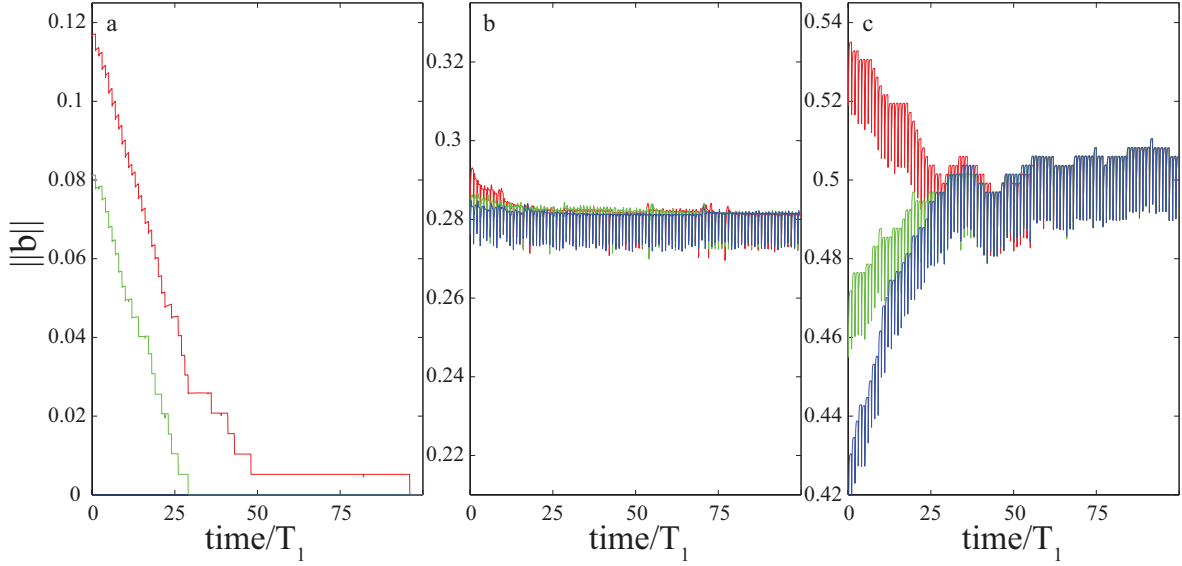}
  %\end{center}
    \caption{Average biomass vs. time, in a system under the influence of periodic local disturbances, for different initial conditions (in red, blue and green).
 \textbf{(a)} For $p=0.4<p_1$, regardless of the initial condition, the number of peaks and, consequently, the average biomass decrease monotonically, so that after a long enough period of time, the bare-soil state is reached.
 \textbf{(b)} For an intermediate precipitation level, $p=0.9$, the only noticeable effect of the disturbance regime is a slight initial decrease of the overall biomass followed by temporary fluctuations.
 \textbf{(c)} For the second bistability range (of nonzero uniform and patterned states), $p=1.3$, the existence of hybrid states dominates the dynamics.
 The system converges to a certain hole density and, with it, to a certain biomass density, regardless of the initial condition.
 Thus, if the initial state is a hybrid solution with many gaps, or a periodic solution with a long wavelength, the overall biomass will grow due to the local disturbance regime.\label{fig:LocDisBio}}
\end{figure}

The response of the system to local disturbances is noticeably different than the response to changes in the global conditions (i.e., changes in $p$).
We looked at the system's response to a regime of local disturbances in three distinct precipitation ranges, the two bistability ranges previously described, and the intermediate precipitation range between them. 
Here, $p$ remains constant during the simulation ($A=0$).
For low precipitation values ($p_L<p<p_1$), the single-peak solution is stable and no splitting occurs. 
In this range, we found that the removal of vegetation spots by local disturbances leads to a redistribution of the remaining spots, with no change in their total number. 
If the system is no longer disturbed, it will asymptotically converge to a patterned state with a longer wavelength. 
However, under continuous repetitions of such local disturbances, the system converges to the bare-soil state, as Fig. \ref{fig:LocDisEvo}a shows (for $p=0.4$).
This result does not depend on the initial conditions, as can be seen in Fig. \ref{fig:LocDisBio}a; the biomass decreases monotonically to zero irrespective of the initial condition, provided that the system is repeatedly disturbed. 

In the intermediate precipitation levels ($p_1<p<p_T$), the single-peak solution does not exist and isolated peaks tend to split.
In this range, a sequence of local disturbances initially modifies the density of the peaks to correspond to the preferred wavelength.
Each consecutive disturbance is followed by a quick response, involving the splitting of peaks, and convergence of the system back to a similar patterned state.
These dynamics are shown in Fig. \ref{fig:LocDisEvo}b for $p=0.9$. 
Graphs of the average biomass versus time for different initial conditions are shown in Fig. \ref{fig:LocDisBio}b. 
The only noticeable effect of the the local disturbances is a slight decrease in the overall biomass.

For higher precipitation rates ($p_T<p<p_S$), the system is highly plastic due to the existence of stable hybrid states.
The hybrid states allow for a multiplicity of stable states, some of which differ only locally.
This allows the system to respond to a local disturbance by changing its state in the immediate vicinity of the disturbed region, without affecting the whole system.
This type of dynamics is shown in Fig. \ref{fig:LocDisEvo}c (for $p=1.3$), where a new hole in the uniform vegetation is created after each disturbance. 
The local response results in a different spatial distribution of the vegetation after each disturbance, 
but gradually the system reaches a hole (vegetation) density in which, on average, for each new hole that is formed, an old one disappears (due to the expansion of vegetated domains), so that the total biomass no longer changes significantly.
Furthermore, the local nature of the response has an additional interesting effect; the initial conditions are completely lost over time, and the state of the system is controlled entirely by the disturbance regime. 

In Fig. \ref{fig:LocDisBio}c, we show the average biomass versus time under a regime of local disturbances. 
The three lines correspond to different initial conditions subjected to the same series of local disturbances. 
It is noticeable that after a large enough number of local disturbances (i.e., a sufficiently long time), all biomass changes are exactly the same for all three different initial conditions shown.
This also means that if the initial conditions are of low overall vegetation, as in a stable periodic solution with a long wavelength or a hybrid state with many gaps,
then the local disturbances actually increase the overall vegetation over time (blue and green curves in Fig. \ref{fig:LocDisBio}c).

%~~~~~~~~~~~~~~~~~~~~~~~~~~~~~~~~~~~~~~~~~~~~~~~~~~~~~~~~~~~~~~~~~~~~~~~~~~~~~~~~~~~~~~~~~~~~~~~~~~~~~~~~~~~~~~~~~~~~~~~~~
%~~~~~~~~~~~~~~~~~~~~~~~~~~~~~~~~~~~~~~~~~~~~~~~~~~~~~~~~~~~~~~~~~~~~~~~~~~~~~~~~~~~~~~~~~~~~~~~~~~~~~~~~~~~~~~~~~~~~~~~~~
\section{Discussion}\label{discuss}
In this paper, we used a simple model describing the spatio-temporal dynamics of water-limited vegetation to study the possible responses of a pattern-forming ecosystem to 
periodically changing environmental conditions (the precipitation rate) and to random local disturbances (biomass removal).
Despite its simplicity, the model offers a complex bifurcation diagram with multistability of patterned states, hybrid states and, in particular, two bistability ranges of patterned states and uniform states. 
For low precipitation rates, we identified a bistability of bare-soil and patterned states, while for higher precipitation rates, we identified a bistability of uniform-vegetation and patterned states. 
In the bistability range of bare-soil and patterned states, we found that there exists a wide range of periodic solutions--from a dense short-wavelength pattern to a sparse long-wavelength pattern (a single biomass peak in the domain studied). 
The bistability range of uniform-vegetation and patterned states shows a smaller range of wavelengths, but compensates for the narrow band of periodic solutions with the existence of multiple stable hybrid states. 
The wide set of hybrid states allows a local response of the system to disturbances and changes in the environmental conditions, and strongly affects the dynamics of the system in this bistability range. 
These differences between the two bistability ranges lead to substantial differences in the system's response to both environmental changes and disturbances.
We note that the results discussed are not limited to a specific model or a specific pattern-forming mechanism. 
We expect the behavior to be general and relevant to systems with a bistability of patterns and uniform states, and more specifically, to systems in which localized states exist inside the bistability range
As shown in the appendix, a qualitatively similar bifurcation structure exists for the Rietkerk et al. model~\cite{Rietkerk2002an}, 
with localized states inside a bistability range of patterns and uniform vegetation (while the bistability between patterns and bare-soil shows no such localized states). 
The similar bifurcation structure implies similar responses of the system to the disturbances and precipitation modulations considered here.

In both bistability ranges, we found that small-amplitude modulations of the precipitation parameter did not result in significant changes in the system's state (red curves in Fig. \ref{fig:PerPerBio}), 
but large-amplitude modulations can result in state transitions that involve a significant biomass reduction (green and blue curves in Fig. \ref{fig:PerPerBio}). 
The manner in which the average biomass is reduced differs between the two bistability ranges. 
In the bistability range of bare-soil and patterned states, large-amplitude precipitation modulations lead to an abrupt collapse to the bare-soil state, 
while moderate amplitude modulations result in a fast but moderate biomass decrease that results from a transition to a sparser periodic pattern.
By contrast, in the bistability range of uniform-vegetation and patterned states, the asymptotic state is always similar--a periodic pattern with a preferred wavelength (provided the modulation is strong enough). 
The approach to this state is typically slower, albeit dependent on the modulation amplitude.

The system's response to repeated local disturbances is also quite different in the two bistability ranges.
In the bistability range of bare-soil and patterned states, the biomass decreases to zero monotonically in time, irrespective of the initial condition (the red, green and blue curves in Fig. \ref{fig:LocDisBio}), 
whereas in the bistability range of uniform-vegetation and patterned states, the biomass dynamics strongly depend on the initial condition. 
In particular, the average (and total) biomass may increase if the initial state comprised a relatively low total biomass.

These differences between the two bistability ranges reveal an aspect of the complex nature of possible transitions in pattern-forming ecosystems.
The model for water-limited vegetation dynamics suggests that the process of desertification may be gradual for higher precipitation rates, but tends to be more abrupt as the system approaches the bare-soil state.
This asymmetry stems from the existence of localized states in the bistability range of uniform vegetation with a patterned vegetation state, which do not occur in the other range. 
The response of similar dryland ecosystems to varying conditions may be more symmetric if localized states occur in both bistability ranges. 
This situation can be expected if both uniform states correspond to nonzero biomass.

Our findings call for further research into systems of higher dimensions in which bistability and multistability regions of patterned states, with different spatial symmetries, may exist.
Furthermore, the response of dryland ecosystems to local disturbances has been less explored than their response to changes in the global conditions. 
Comparisons of model predictions with field observations of ecosystem dynamics under different regimes of local disturbances can enhance our understanding of these ecosystems and improve the mathematical models describing them.

\setcounter{equation}{0}
\appendix*
\section{Appendix}
In the main text, we considered a simplified version of the Gilad et al. model \cite{Gilad2004prl,Gilad2007jtb,zelnik2013regime}, which captures only the ``uptake-diffusion`` pattern-forming mechanism. In order to demonstrate that our results are not limited to this specific mechanism (or model), we investigate here the bifurcation diagram of a different model capturing the ''infiltration contrast`` mechanism. 
The results presented in the main text stem from the existence of two bistability ranges of periodic patterns and uniform states. 
The existence of localized states within one of the bistability ranges strongly affects the response of the system to disturbances.
Therefore, we consider the existence of similar bifurcation diagrams (including the abovementioned two bistability ranges and localized states within only one of these ranges) in different models to be imperative for the generality of our results. 
We consider here a model by Rietkerk et al. \cite{Rietkerk2002an, HilleRisLambers2001ecology}. The model describes the dynamics of belowground ($W$) and aboveground ($h$) water and the aboveground biomass ($b$). 
This three-variable model consists of the following non-dimensional equations \cite{zelnik2013regime}:
\begin{align}
 \partial_t b&=G(w) b-\mu b+\nabla^2b\,, \label{Rb} \\
 \partial_t w&= Ih - \nu w -\gamma G(w) b +D_w\nabla^2 w\,, \label{Rw} \\
 \partial_t h&=p- Ih +D_h\nabla^2h\,, \label{Rh}
\end{align}
where
\begin{equation}
    G=\frac{w}{w+1}\,,\qquad I=\alpha\frac{b+f}{b+1}\,.\label{GI}
\end{equation}
In equation \eqref{Rb}, the biomass growth rate, $G=G(w)$, depends on the soil water variable only; the dependence is linear at small soil water contents and approaches a constant value at high contents, representing full plant turgor. 
Biomass growth is also affected by mortality ($-\mu b$) and by seed dispersal or clonal growth ($\nabla^2 b$). 
The soil water content (equation \eqref{Rw}) is increased by the infiltration of surface water ($Ih$). 
The biomass dependence of the infiltration rate, $I=I(b)$, captures the infiltration contrast that exists between bare soil (low infiltration rate) and vegetated soil (high infiltration rate) for $f<1$.  
The other terms affecting the dynamics of the soil water represent the loss of water due to evaporation and drainage ($-\nu w$), water uptake by the plants ($-\gamma Gb$), and moisture diffusion within the soil.
The surface water dynamics (equation \eqref{Rh}) are affected by precipitation at a rate $p$, by water infiltration into the soil, and by overland flow modeled as a diffusion process.

This model captures a pattern-forming mechanism: the infiltration feedback. When the infiltration contrast is high ($f\ll 1$), patches with dense vegetation act as sinks for runoff water. 
This accelerates the vegetation growth, sharpens the infiltration contrast and increases, even further, the soil water content in the vegetated areas. 
The water flow towards vegetation patches inhibits the growth in the patch surroundings, thereby promoting vegetation pattern formation. The model is expected to describe the dynamics of water-limited vegetation in areas where soil crust tends to form. The vegetation cracks the crust, thereby increasing the infiltration rate. 

\begin{figure}
\includegraphics[width=\linewidth]{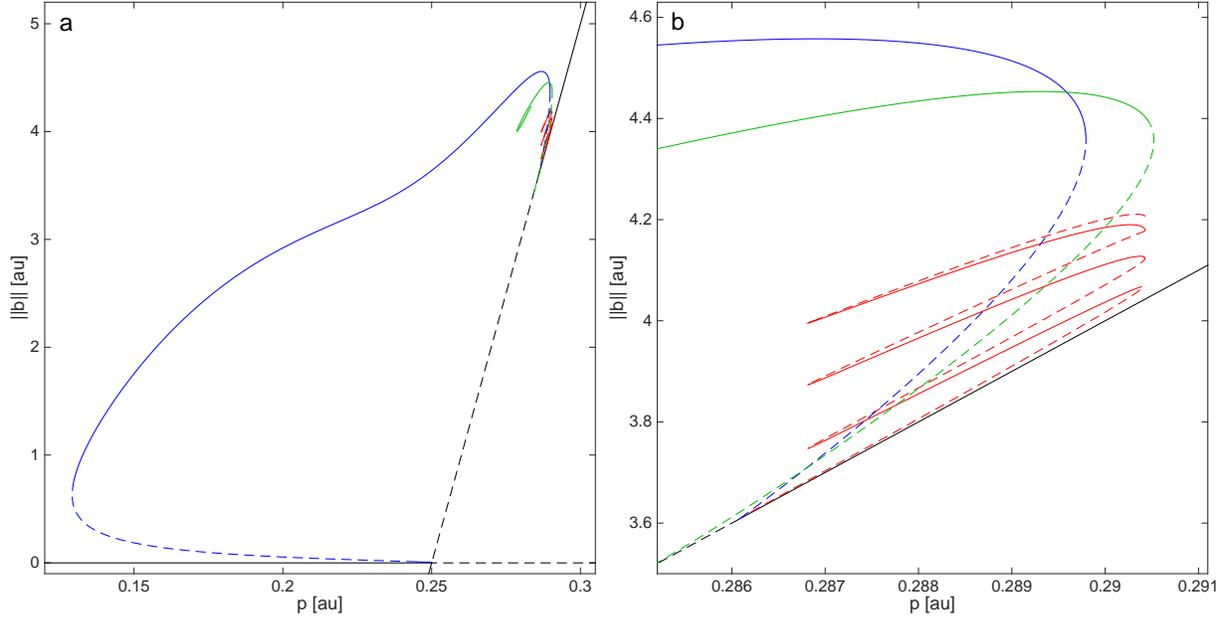}
\caption{
Bifurcation diagrams of the Rietkerk et al. model in one spatial dimension. 
\textbf{a:} A bifurcation diagram showing the two ranges of bistability between uniform states and patterned states. 
\textbf{b:} A blowup showing the bistability range between patterns and uniform vegetation, including the localized states. 
The vertical axis denotes the L2 norm of the biomass density while the horizontal axis denotes the precipitation rate $p$. 
Solid (dashed) curves denote stable (unstable) solutions. The shown steady states include the two uniform states of the system 
(black curves, corresponding to the bare-soil ($b=0$) and uniform-vegetation states), the Turing patterned state (green curve), 
and a periodic solution that extends to the boundary of the bistability range. 
The model parameters are: $\mu = 0.20$, $\alpha = 0.50$, $f = 0.2$, $\ni = 1.0$, $\gamma = 0.05$, $D_w = 1$, $D_h = 1000$.
\label{fig:Rsnaking}
}
\end{figure}

For certain parameter ranges and relatively high precipitation rates, the model shows a bistability of patterned states and uniform-vegetation states, as well as stable localized states within this bistability range. For lower precipitation rates, the model shows a bistability of patterned states and the bare-soil state without stable localized states.
A bifurcation graph that includes these localized states is shown in Fig. \ref{fig:Rsnaking}.

%\section{}

\begin{acknowledgments}
The research leading to these results has received funding from the European Union Seventh Framework Programme (FP7/2007-2013) under grant number [293825], and from the Israel Science Foundation under grant number~305/13.
\end{acknowledgments}

\bibliography{CP4.bib}

\end{document}